# The continuum limit of the lattice Gribov problem, and a solution based on Hodge decomposition.

Philippe de Forcrand [a] and James E. Hetrick [b]

[a]Interdisciplinary Project Center for Supercomputing (IPS)
ETH-Zentrum, CH-8092 Zürich, Switzerland

[b]Physics Dept., University of Arizona
Tucson, AZ 85721, USA

We study gauge fixing via the standard local extremization algorithm for 2-dimensional $U(1)$. On a lattice with spherical topology $S^2$ where all copies are lattice artifacts, we find that the number of these 'Gribov' copies diverges in the continuum limit. On a torus, we show that lattice artifacts can lead to the wrong evaluation of the gauge-invariant correlation length, when measured via a gauge-fixed procedure; this bias does not disappear in the continuum limit. We then present a new global approach, based on Hodge decomposition of the gauge field, which produces a unique smooth field in Landau gauge, and is economically powered by the FFT. We also discuss the use of this method for examining topological objects, and its extensions to non-abelian gauge fields.

Gauge fixing to a smooth (Coulomb or Landau) gauge on the lattice is essential for the extraction of non-perturbative, gauge-dependent observables. It also allows a direct comparison of non-perturbative results with continuum, perturbative ones. Finally it is emerging as a powerful approach to measuring gauge-invariant observables: like blocking or cooling, it removes the noise of ultra-violet fluctuations; but unlike these, it preserves the Yang-Mills action. For all these reasons, one is interested in generating smooth, continuum-like gauge fields on the lattice. However the faithful representation of continuum gauge fields is a subtle issue, complicated both by the compact formulation and the lattice representation of the underlying space-time which allows violations of cherished geometric identities. For instance, the compact formulation allows deviations of the Bianchi identity $dF = d^2 A = 0$ which introduce monopoles and cause lattice QED to confine [1]. Furthermore, the lattice representation of a fiber bundle with the gauge group attached to sites, apparently has quite a different de Rham cohomology than the continuum version. This is responsible for the 'lattice Gribov problem' in which gauge fixing algorithms find a huge number of configurations which satisfy the gauge condition, introducing Aharonov-Bohm type objects such as Dirac sheets, strings, and vortices. How then do we sample over the set of copies, of continuum or lattice origin, and what is the proper weight ?

## 1. Preliminaries

To exhibit the deviations of the lattice gauge field topology from the continuum which occur in gauge fixing, we use the concise geometric notation of differential forms and here review some definitions [3]. A fundamentally useful relation is Hodge Decomposition which states that every $r$–form $A^{(r)}$ (in an $n$-dimensional manifold $M$) is *uniquely* decomposed into

$$A^{(r)} = \delta\chi^{(r+1)} + d\phi^{(r-1)} + h^{(r)} \qquad (1)$$

where $d$ is the exterior derivative, $\delta = (-)^{nr+n+1} * d*$ is its dual via the Hodge star $*$, and $h^{(r)}$ is a harmonic $r$–form that satisfies $dh = \delta h = 0$. For a vector field in 3 dimensions, (1) reads in coordinates:

$$A_i = \epsilon_{ijk}\nabla_j \chi^k + \nabla_i\phi + h_i \qquad (2)$$

A more technical point also due to Hodge is the relationship

$$\dim\{h^{(r)}\} = \dim H^r(M) = b_r \qquad (3)$$

between the harmonic forms $\{h^{(r)}\}$ and the de Rham cohomology groups $H^r$ (and hence the Betti numbers $b_r$). Recall that $H^r$ classifies, roughly, the non-contractible $r$-spheres in $M$ and hence its (global) topology, while $\{h^{(r)}\}$ are the $r$-dimensional zero eigenvectors of the Laplacian, a (local) differential operator.

### 1.1. Existence of copies in the continuum

For a given gauge group $G$ and (compact, Euclidean) spacetime manifold $M$, we know when to expect generic Gribov copies [4,5]. Consider $M = S^{n_1} \times S^{n_2} \times ... \times S^{n_k}$, a direct product of spheres of various dimensions, which includes the usual choice of a 4-torus. Then a sufficient condition for the existence of Gribov copies is that an integer $m$ can be found in $\{1, 2, ..., \sum n_k\}$ such that the homotopy group $\pi_m(G)$ is non-trivial. This theorem arises because gauge transformations map the de Rham cohomology structure of $M$ (which is rather simple for spheres) onto the gauge group.

For example, all $SU(N)$ groups have $\pi_3(G) = Z$, and thus Gribov copies will be found on any sphere or torus of $d \geq 3$, including the original case studied by Gribov [6]: Coulomb gauge on the 3-sphere.

For $G = U(1)$, $\pi_1(G) = Z$ and thus one expects copies on any torus ($S^1 \times anything$), but not on a sphere. The difference is clear: a closed Polyakov loop around a torus may have non-trivial winding; on a sphere such a loop is contractible and is unwound in the process of gauge-fixing.

### 1.2. Lattice artifacts

On the lattice additional gauge copies exist besides the continuum copies considered above. In addition to the large scale (continuum) topology of the manifold $M$, there is a local structure due to the elementary lattice hypercubes (plaquette, cube, or 4-d hypercube), which allows for non-contractible gauge transformations, and gives rise to gauge ambiguities.

These ambiguities are localized lattice artifacts, like XY-vortices, Dirac strings, etc... They appear because the gauge transformation between two lattice 'Gribov' copies generically maps the boundary of some hypercubes onto $G$ with non-trivial winding [2], and is also a zero mode of the lattice Laplacian making it an element of $\{h^{(r)}\}$. Thus when we construct a lattice fiber bundle by attaching a group manifold to each site, the underlying cohomology of the spacetime appears as one with holes at every hypercube, ie. dim $H^r$ = dim $\{h^{(r)}\}$ diverges in the continuum limit for certain $r$. More accurately it is the homotopy groups $\pi_r(M)$ of the fiber bundle which classify gauge transformations that are affected.

Since they do not induce any net global topology (otherwise they would belong to continuum copies), these lattice artifacts must come by pairs with opposite winding, ie. vortex pairs. The gauge transformation which creates a vortex pair creates the analog of an Aharonov-Bohm flux tube, whose gauge field is harmonic. However, unlike a continuum flux tube, such a gauge transformation on the lattice can be continuously deformed into the identity.

Including these spurious copies in the partition function is then overcounting; moreover observables may take different values on such copies. Thus it is crucial to control the occurrence of vortex pairs and any bias they may cause.

## 2. Numerical gauge fixing by the standard (local) method

The standard approach, in essence, is to ignore the Gribov problem: gauge-fixing is performed by a local iterative procedure which fails to eliminate lattice artifacts, and each gauge-fixed configuration is given the same weight. The hope is that this approach introduces no bias, and converges to the correct result in the continuum limit.

### 2.1. $U(1)$ on $S^2$: a study of lattice artifacts

We have examined this expectation for the standard choice of lattice Landau gauge

$$f(G, \{U\}) = \frac{1}{N} \sum_{i=1}^{N} ReTr U_i^G \Big|_{\max} \qquad (4)$$

and for the standard local iterative methods of maximizing $f$ (relaxation and over-relaxation

Figure 1. Vortex pair distribution for cubes of size 6 to 12.

of point singularities, compared to its continuum gauge-fixed counterpart.

Notice that this conclusion could not be reached if one would simply monitor the trace defect $\Delta$. As Table I indicates, $<\Delta>$ seems to slowly decrease to zero as $\beta \to \infty$, misleading one toward optimism. In fact, $f(G,\{U\})$ is the Hamiltonian of a spin-glass with couplings $U$, and the situation we encounter is typical of the search for a ground-state, eg. by simulated annealing: as the system size increases, the energy per site approaches that of the true ground state, even as the distance in phase space between the 2 states diverges.

| cube size | 6 | 8 | 10 | 12 |
|---|---|---|---|---|
| $\beta$ | 5 | 8.89 | 13.89 | 20 |
| <# vortex pairs> | 0.65 | 0.99 | 1.09 | 1.33 |
| <trace defect> | 0.017 | 0.017 | 0.014 | 0.013 |

Table 1
Evolution with $\beta$ of the average properties of gauge copies.

To understand better the behavior of $\Delta$, we looked at its correlation with the number of vortex pairs. Within our statistics, $\Delta$ grew in proportion to the number of pairs, confirming our expectation that vortices form a dilute, non-interacting gas. Further evidence was provided by the distribution of the mutual angle, on the sphere, between a vortex and an anti-vortex: it was uniform, except for a sharp drop to zero at small angles, corresponding to $\sim 2$ lattice spacings. This uniform distribution reflects the failure of the local maximization used, to bring together and annihilate vortex and anti-vortex unless they are next to each other on the lattice.

The trace defect per vortex pair can be approximately calculated by considering the prototypical lattice copy on the sphere: a vortex at the North pole and anti-vortex at the South pole, as if created by a Dirac string piercing the sphere along the polar axis. The harmonic gauge field produced by the string is $A = d\phi/r\sin\theta$, and the trace defect is $\frac{1}{\#links}\int_{S^2} A \wedge *A$. Taking for the number of links $8\pi R^2/a^2$, and integrating over the whole surface of the sphere except for 2 circular caps of radius $a$, one gets

$$\Delta \sim -\frac{1}{4}\frac{a^2}{R^2}\log\frac{a^2}{R^2} \qquad (5)$$

which goes to zero with $a$. In actuality the vortices are not systematically at opposite poles, and the integration is really over a discrete cube, so that the coefficients in (5) will be changed. Still a formula of the type (5) describes our data well, as Fig.2 shows.

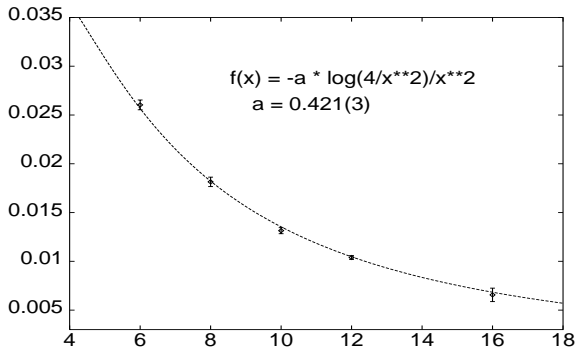

Figure 2. Trace defect per vortex pair, as a function of the lattice spacing.

This simple toy model shows the fatal disease of the standard local approach to gauge fixing: **as the lattice spacing decreases, gauge-fixed configurations have less and less to do with their continuum counterpart**. We do not expect this conclusion to change with the gauge group or the lattice manifold.

### 2.2. Measurable effects of lattice artifacts

Several attempts have been made at detecting observable effects of spurious gauge copies [8–10]. This is a very difficult task, since one does not know how to systematically identify and eliminate these copies. At best, one can establish a correlation between the value of some physical observable and the "quality" of gauge fixing [11,12]. It also seems to depend on the sensitivity of the observable to topology.

We decided to look at this issue again on a toy model, $U(1)$ on a 2-dimensional torus. Specifically we tried to extract the correlation length $\xi$ of 2 Polyakov loops. In this model each link $U$ is defined by an angle $\theta_\mu(x) = \frac{1}{i}\log U_\mu(x) \in [-\pi,\pi]$. We measured this correlation in 3 different ways:

**(1)** $< e^{i(\Theta_P(t)-\Theta_P(0))} >$ This is the usual gauge-invariant procedure, where $\Theta_P(t) = \sum_{x=1}^L \theta_x(x,t)$. We repeated the measurement for 3 values of $\beta$, from 16 to 256, on $16^2$ to $64^2$ tori, keeping the physical size constant. The lattice model is exactly solvable in this case [13], providing a check of our Monte Carlo program.

**(2)** $< (\Theta_P(t)-\Theta_P(0))^2 >$. This is the correlation one would measure after fixing to the nearest maximum of $f$ in Coulomb gauge, with $\theta_x(x,t) = 1/L\ \Theta_P(t)$.

**(3)** $< (\tilde\Theta_P(t)-\tilde\Theta_P(0))^2 >$ where $\tilde\Theta$ is the projection of $\Theta$ in $[-\pi,\pi]$. This corresponds to fixing to Coulomb gauge and selecting the global maximum of $f$.

One can see in Fig.3 that **(3)** is a noisy but correct way to measure the correlation length $\xi$, but that **(2)** is completely wrong. In addition, there is no trend for **(2)** to approach the correct result as $a$ decreases. Thus we have exhibited one situation where lattice copies completely spoil a physical measurement, and appear to continue to do so in the continuum limit.

## 3. Gauge fixing by Hodge decomposition: a global approach

### 3.1. Abelian gauge fields

Since we have shown above that the lattice Gribov problem is essentially due to lattice artifact harmonic forms, Hodge decomposition provides us with a very powerful tool for separating the continuum gauge fixed fields from the lattice defects. While the exterior algebra on a lattice [14] is defined for non-compact complex fields, we extend it to an abelian lattice fiber bundle by using link angles $\theta_\mu(x) \equiv A_\mu(x)$. From (1) we have

$$dA \quad = F_{\mu\nu} = \Delta\chi \qquad (6)$$
$$\delta A \quad = \partial_\mu A_\mu = \Delta\phi. \qquad (7)$$

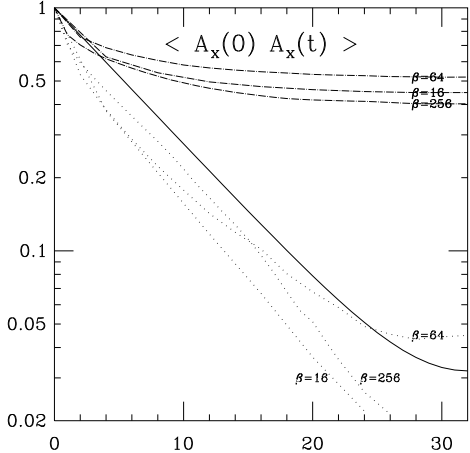

Figure 3. Gauge-fixed correlations on the torus, with (dashed line) and without (dotted line) spurious gauge copies. The solid line is the gauge-invariant analytic result.

Here $\Delta = d\delta = \delta d$ is the lattice Laplacian ($\Delta_{x'x} = \sum_\mu \delta_{x'x+\mu} - 2\delta_{x'x} + \delta_{x'x-\mu}$).

The Landau gauge condition is $\delta A = 0$, which when translated into links is

$$\delta A = 0 \Rightarrow \prod_\mu U_\mu^\dagger(x-\hat\mu)U_\mu(x) = 1 \quad \forall x. \qquad (8)$$

We take this as the definition of our Landau gauge condition for abelian fields, and will address non-abelian fields shortly.

There are 2 possible methods to fix to Landau gauge then. The first one is to compute $\delta A(x) = \sum_\mu [A_\mu(x) - A_\mu(x-\hat\mu)]$. From this compute $\phi(x)$ in the Hodge decomposition of $A$ by inverting the Laplacian in (7) (via the Fast Fourier Transform). From $\phi$, make the gauge transformation

$$A \to A' = A - d\phi = \delta\chi + h. \qquad (9)$$

$A'$ now satisfies $\delta A' = 0$ identically, but is not unique since $h$ may be any one of the lattice artifact harmonic forms.

We could proceed and eliminate the harmonic part $h$ by finding the gauge transformation corresponding to the topological object described by $h$ (eg. vortex pairs in 2-D), however this is cumbersome. It is far simpler to just set

$$A \to A' = \delta\chi = \delta(\Delta^{-1}F). \qquad (10)$$

We display this method in Figs.4-6 ($L=16$). The initial field $U_1(\vec x) = \frac{\pi}{2}\sin(2\pi x/L)\cos(2\pi y/L)$, $U_2(\vec x) = 1$, was transformed with a random gauge transformation (Fig. 4). The gauge fixed configuration is shown in Fig. 5. For comparison, we show also the same field gauge fixed by the local method of section 3 in Fig. 6.

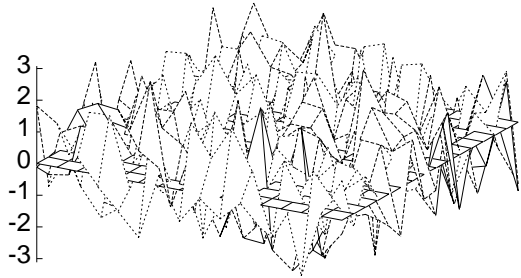

Figure 4. Initial gauge field configuration.

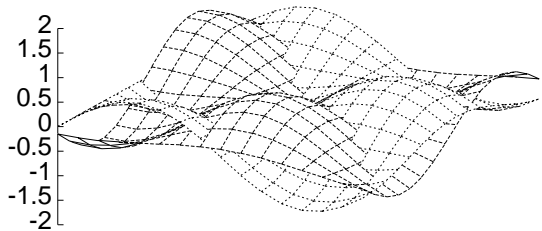

Figure 5. Landau gauge fixed field configuration: Hodge method $A' = \delta(\Delta^{-1}F)$.

Notice that we have some freedom in computing $F$ in (10) which is very interesting. If we use the compact definition of $F = \frac{1}{i}\log U_{\text{plq}}$, the gauge fixed field $A'$ will preserve the DeGrand-Toussaint monopoles which violate $dF = 0$. Using the non-compact definition of $F_{\mu\nu} = A_\nu(x+\hat\mu) - A_\nu(x) - A_\mu(x+\nu) + A_\mu(x)$, will set $A'$ to the equivalent non-compact gauge field configuration, *without* lattice monopoles. Thus we have a way of easily converting a field from compact form to non-compact form, preserving the action, and gauge fixing at the same time.

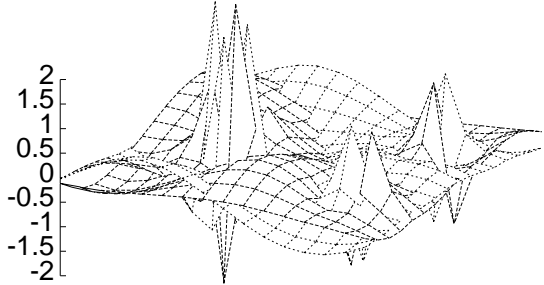

Figure 6. Gauge fixed field configuration: Standard maximization method of section 3

### 3.2. Non-abelian gauge fields

The above gauge fixing method works perfectly for abelian fields since the discrete exterior algebra lifts (up to compactness issues) nicely to the abelian fiber bundle. However, the situation is rather complicated for non-abelian fields. We still have Hodge decomposition (1) for Lie algebra valued $r$-forms $A_a^{(r)}\tau^a$, however $dA$ is not a gauge invariant quantity, since it is the covariant exterior derivative of $A$, $DA = F$ which is the field strength. Furthermore, nilpotence is lost for $D$ on Lie algebra valued forms: $D^2\Phi = F \wedge \Phi$. Thus the program above must be modified.

We found interestingly, that if we apply the first method outlined for the abelian case, ie. extract $\phi^a = \Delta^{-1}(\delta A^a)$ and make the gauge transformation $U \to G^\dagger(x)U_\mu(x)G(x+\hat\mu)$, where $a$ is the $SU(2)$ algebra index and $G = \exp(-i\epsilon\phi)$ that, for $0 < \epsilon < 1.3$, successive gauge transformations iteratively converge to $\delta A^a = 0$.

There are a variety of geometrical extensions of this method which we are exploring. We present next a simple, smooth, and unique gauge for non-abelian fields.

### 3.3. The $\tau$ gauge

Since we have a unique gauge fixing method for abelian fields we propose the following algorithm for gauge fixing non-abelian fields.

First, choose an operator $O(x)$ which transforms equivariantly ($O^G(x) = G^\dagger(x)O(x)G(x)$) and can thus be rotated at each lattice site to a fixed direction in '$\tau$' space by a gauge transformation. This leaves only an abelian gauge symmetry corresponding to transformations which commute with the chosen vector in $\tau$ space. Then, following the above method based on Hodge decomposition, gauge fix this remaining abelian freedom.

For an $SU(2)$ example, take $O$ to be the sum of plaquettes at each site, projected onto SU(2), and diagonalize it. We are left free to make gauge transformations $G(x) = \exp[i\alpha(x)\tau_3]$ which we can use to set, say, $\delta A^3 = 0$. Thus this gauge has a continuum limit of

$$\sum_{\mu<\nu} F_{\mu\nu}^{1,2} = 0, \quad \partial_\mu A_\mu^3 = 0 \qquad (11)$$

This gauge could be a simple and useful choice, especially for the study of monopole-induced confinement since it is very much like t'Hooft's original proposal. It is economical, since the global step can be performed by FFT. It is unique, up to known continuum harmonic forms. Furthermore it has a convenient perturbative interpretation, unlike earlier global proposals [15].

### REFERENCES

1. T.A. DeGrand and Doug Toussaint, *Phys. Rev.* D22 (1980) 2478; T. Banks et al., *Nucl. Phys.* B129 (1977) 493.
2. Ph. de Forcrand et al. *Nucl. Phys.*, B20 (Proc. Suppl. 1991) 194.
3. M. Nakahara, *Geometry, Topology, and Physics*, Adam Hilger Pub. 1990.
4. I.M.Singer, *Comm.Math.Phys.*, 60 (1978) 7.
5. T.P.Killingback, *Phys.Lett.*, B138 (1984) 87.
6. V.N.Gribov, *Nucl.Phys.*, B139 (1978) 1.
7. C.B. Lang and T. Neuhaus, *Nucl. Phys.* B34 (Proc. Suppl. 1994) 543.
8. A.Nakamura and M.Plewnia, *Phys. Lett.* B255 (1991) 274.
9. S. Hioki et al., *Phys. Lett.* B271 (1991) 201.
10. M.L. Paciello et al., preprint ROME-1034-1994,`hep-lat 940912`.
11. V.G. Bornyakov et al., *Phys. Lett.* B317 (1993) 596; and these proceedings.
12. V.G. Bornyakov et al., these proceedings.
13. B. Rusakov, *Mod. Phys. Lett.* A5 (1990) 693.
14. H. Joos, *Zeit. Phys.* C15 (1982) 343.
15. J. C. Vink and U-J Wiese,*Phys. Lett.* B289 (1992) 122; P. van Baal, these proceedings.